# FILE SYNCHRONIZATION SYSTEMS SURVEY

Zulqarnain Mehdi[1] and Hani Ragab-Hassen[2]

[1]School of Mathematical and Computer Sciences, Heriot-Watt University
zm82@hw.ac.uk
[2] School of Mathematical and Computer Sciences, Heriot-Watt University
h.ragabhassen@hw.ac.uk

## ABSTRACT

*Several solutions exist for file storage, sharing, and synchronization. Many of them involve a central server, or a collection of servers, that either store the files, or act as a gateway for them to be shared. Some systems take a decentralized approach, wherein interconnected users form a peer-to-peer (P2P) network, and partake in the sharing process: they share the files they possess with others, and can obtain the files owned by other peers.*

*In this paper, we survey various technologies, both cloud-based and P2P-based, that users use to synchronize their files across the network, and discuss their strengths and weaknesses.*

## 1. INTRODUCTION

Sharing digital files over a network is a common application of the networking technology. Files can be shared between: a) Users and machines (eg: when one downloads a file from a server, or uploads a file to a server), b) Machines (eg: automated backups), c) Different users (through machines: uploading the file to a server, from which the other party can download it; directly: using P2P file sharing services). It is common nowadays for users to share files across their own devices connected over a network using synchronization services such as Dropbox [1] or Google Drive [2]. They usually do so by allowing a user to upload their files from one device to central servers, and allowing other devices owned by the same user to download them from those servers. Note that the users also have the option to share their files with others, or make them public. Peer-to-peer (P2P) based synchronization systems split the files into chunks (or pieces), which are then replicated on a subset of peers. The original peer's other devices can retrieve the chunks and combine them to form the original file. In this paper we review the main existing file synchronization systems, and compare them. The rest of this paper is organized as follows: section II introduces important backbone technologies. Notable existing file synchronization systems are reviewed in section III. We discuss those system in section IV. Section V concludes our paper.

# 2. BACKBONE TECHNOLOGIES

Prior to diving deeper into the survey, it is important to have some knowledge of the underlying technologies of the reviewed services. While the technologies used by cloud-based storage systems are quite straight-forward, P2P infrastructures are of a more complex nature. This is why we briefly describe important P2P protocols in this section.

It should be noted that P2P might have slightly different meanings in different contexts. The definition presented in this paper is according to [3]. In a P2P system, each peer in the system provides the service that it is intended to, by sharing its resources (eg: storage and processing power). These peers communicate directly, without the need of an intermediate node.

A pure P2P system is fully decentralized, but partially centralized P2P systems do exist. The best example of such a P2P system is the BitTorrent protocol, wherein a central server, a tracker, tracks the peers currently downloading each file. Other peers can then contact the tracker and request the list of these peers, and contact them.

For a truly decentralized P2P network to exist, the nodes first need to find other nodes in the network (peer discovery). In a local network, a simple scan could reveal other nodes that participate, or are interested in participating in a local P2P network. Over a wider network (such as the Internet), however, this would be a non-trivial task, as it would be unfeasible for a node to look up the entire network to connect to the ones that share the same interests.

There are a few protocols that allow peers to discover each other in a P2P network. In the following subsection, we review Pastry [4], a P2P discovery protocol; we then review BitTorrent, the defacto P2P standard.

## 2.1. Pastry

In [4], Rowstron and Druschel presented Pastry a new object location protocol for large scale P2P systems. Pastry performs application level node look up and routing over a large network connected via the Internet; when a node receives a message along with a key, it routes the message along all the live nodes to the node which has a nodeID 'numerically' closest to the key. Each node in Pastry keeps track of its immediate neighbors, and notifies other nodes of any changes in the network, such as when a new node joins the network, or if one leaves the network.

Pastry is completely decentralized, and aims to reduce the routing steps that messages have to take to reach the destination. The expected number of routing steps in Pastry is $O(\log N)$, where N is the number of Pastry nodes in the network.

A nodeID is randomly calculated, and ranges from 0 to $2^{128}-1$, allowing them to be "diverse in geography, ownership, jurisdiction, etc." A node is said to be "close" to be another node if its nodeID is numerically close to the key that it receives along with the message. The message is routed to one of such closest nodes in Pastry, which is usually a node near the originator node.

An example of an application of Pastry is PAST [5], a largescale P2P file storage utility, developed by the same authors. More on PAST is detailed in the next section.

## 2.2. BitTorrent

According to the official specification, [6], BitTorrent is a P2P file sharing protocol, used to transfer files of any size across the web, and according to [7], it was created by Bram Cohen to replace the standard FTP. It uses a server (called a tracker) that tracks the files, and aids the clients in downloading and combining the chunks (pieces, according to [6]) of the file into the original file. There are, however, "trackerless" implementations of the protocol, which create a true decentralized environment for BitTorrent based P2P file transfers.
Unlike a typical P2P network, BitTorrent ensures that each client uploads files while downloading files from other peers, ensuring fairness, better availability of files, and a boost in performance.

# 3. FILE SYNCHRONIZATION SYSTEMS

This section presents the most notable file synchronization systems. We distinguish two major categories: cloud-based file synchronization systems, and P2P-based file synchronization systems.

## 3.1. Cloud-based File Synchronization Systems

A cloud-based synchronization system (also a cloud-based storage service) is used to store users' files in a central server, owned and governed by a certain entity (eg: an enterprise, or a small company). Users upload their files to this server from one device, and download them on another (or on the same device, in case the user loses the original file). Users can also share their files with others, and depending on the service provided, a cloud-based synchronization service can be extended to provide a collaboration platform to the users.

These services are provided across many different platforms, using web as well as native application development technologies as their front-end. Some of them provide desktop applications that act as drives connected to the PC, to provide a seamless interaction with the actual cloud drive. These services usually employ a freemium model: a fixed amount of initial storage is given for free, with limited feature set, while allowing users to upgrade to a higher plan with more storage and additional features. A good comparison of some of the most popular cloud storage and synchronization services can be seen in [8]. Such a model makes cloud services much more accessible and convenient to the users.

### 3.1.1. Google Drive

Google Drive [2] is a file storage and synchronization service by Google. At the time of writing this paper, new users to the service get 15 GB of storage for free, with various monthly subscription plans available for more storage [9].

Users can not only store and synchronize their files using Google Drive, they can also view, modify, delete, and in some instances, collaborate on them with other users, using either the web interface, or a native applications available on major platforms. Google Drive supports a plethora of file formats for a user to store, synchronize, and work with.

### 3.1.2. OneDrive

OneDrive [10] by Microsoft is a file storage and synchronization service with similar features to Google Drive, and is powered by Microsoft Azure [11], Microsoft's cloud computing platform. As of January 2016, OneDrive has dropped down its storage capacity for new users from 15 GB to 5 GB. Users who had obtained the 15 GB previously would retain it. Like Google Drive, OneDrive allows users to upgrade the storage using one of the various monthly subscription plans [12].

Along with file storage and synchronization, OneDrive allows users to view, update and delete the files, and collaborate on them using Office Online - a free online Microsoft Office utility.

### 3.1.3. iCloud Drive

A cloud storage and synchronization service identical to Google Drive and OneDrive, iCloud Drive [13] by Apple offers similar features to users as the previously mentioned cloud-based services. In terms of file storage capacity, iCloud Drive offers 5 GB of free space to new users, like Microsoft's OneDrive, with plans for upgrade available [14].

According to [15], iCloud is utilizes both Amazon Web Services (AWS) by Amazon [16], and Microsoft Azure [11] since 2011 (when iCloud first launched). However, there are numerous reports which state that Apple is siding with Google's Google Cloud Platform [17] to provide some of iCloud's services [18] [19] [20].

### 3.1.4. Dropbox

Dropbox [1] is one of the most popular file storage and synchronization service, created not by large entities such as those mentioned above, but by a startup company of the same name.

Dropbox offers 2 GB of storage space initially to new users, with options to upgrade to 1 TB, with a monthly subscription (or unlimited storage for Business users) [21].

## 3.2. P2P-based File Synchronization Systems

A P2P-based synchronization system, unlike a cloud-based synchronization system, is a decentralized system wherein each peer in the network acts as both a server, as well as a client, to synchronize files between a user's authorized devices. In this system, files are broken down into encrypted pieces, and each peer uploads a certain number of pieces to, and downloads from, other nodes in the system, ensuring that the files are almost always available for synchronization, and that no one peer contains the complete file, thus enforcing privacy and security of the users' data. Furthermore, the load is divided among the connected peers, rather than a single server, thus increasing the performance of the synchronization process.

Like centralized cloud synchronization services, P2P service providers provide a similar business model of a free though limited plan, while setting additional storage space up for purchase. However, unlike centralized cloud storage and synchronization services, it is much more efficient and convenient to conjure a private P2P cloud service with possibly unlimited storage (as storage space depends upon the storage shared by each node many nodes equal a lot of storage).

Below are some of the examples of such a system.

### 3.2.1. PAST

PAST is an application of Pastry, developed by the developers of Pastry themselves. PAST extends Pastry's capabilities to form a peer-to-peer file storage system that uses a file's name, as well as the owner's name, to calculate a hash which is used as its fileID. The fileID is used as the key in PAST.

### 3.2.2. Symform

Symform by Quantum [22] is a popular P2Pbased file synchronization service, in which each node forms a cloud in the decentralized network, and contributes its resources (storage space), while receiving certain amount of space itself from other nodes.

In Symform, files are broken down into blocks, encrypted, and spread across the network. This way, the files are always available for synchronization, privacy is maintained, security is enforced, and the synchronization performance enhanced on the network.

### 3.2.3. Resilio Connect

Resilio Connect (formerly Sync, by BitTorrent, Inc.) [23] creates a P2P cloud using BitTorrent among a user's devices, rather than including external nodes into the network. This makes the cloud even more secure, but reduces the reliability of the synchronization service, as offline nodes cannot transmit or receive files.

## 4. DISCUSSION

Table 1 compares the existing technologies and services we mentioned in the previous section. As can be seen from the table, P2P-based file synchronization systems tend to offer the most value to the consumers than the cloud-based services in terms of storage capacity.

Table 1. Comparison of existing file synchronization technologies and services.

|  | Google Drive | OneDrive | iCloud | Dropbox | Symform | Resilio Connect | PAST |
|---|---|---|---|---|---|---|---|
| Free initial storage (GB) | 15 | 5 | 5 | 2 | 10 | Unlimited | Unlimited |
| Cheapest storage plan | 100 GB ($1.99/month) | 50 GB ($1.99/month) | 50 GB* | 1 TB ($9.99/month) | 100 GB ($10/month) | Unlimited storage for free | Unlimited for free |
| Maximum storage | 30 TB | 1 TB (with Office 365) | 1 TB | 1 TB | 1 TB | Unlimited storage | Unlimited storage |
| Platform support | Windows, OSX, Android, iOS, Web | Windows, OSX, Android, iOS, Web | Windows, OSX, Android, iOS, Web | Windows, OSX, Linux, Android, iOS, Web | Windows, OSX, Linux, Android, iOS | Windows, OSX, Linux, Android, iOS | Not mentioned (potentially every platform) |
| Cloud-based or P2P-based | Cloud-based | Cloud-based | Cloud-based | Cloud-based | P2P-based | P2P-based (BitTorrent) | P2P-based |
| Online collaboration | Yes | Yes | Yes | Yes | No | No | No |
| File storage before synchronization | Files are stored on the cloud drive (and on the local drive of the device that uploaded the file, if not deleted) | Files are stored on the cloud drive (and on the local drive of the device that uploaded the file, if not deleted) | Files are stored on the cloud drive (and on the local drive of the device that uploaded the file, if not deleted) | Files are stored on the cloud drive (and on the local drive of the device that uploaded the file, if not deleted) | Files are transferred from the device to the P2P cloud formed by Symform, and stored there as long as the user wishes | Files are stored in the devices where they are first added, and transferred only when a sibling device comes online | Files are transferred from the device to the P2P cloud formed by PAST, and stored there as long as the user wishes |
| File storage after synchronization | Files stay in the cloud drive for as long as the user wishes | Files stay in the cloud drive for as long as the user wishes | Files stay in the cloud drive for as long as the user wishes | Files stay in the cloud drive for as long as the user wishes | The pieces of the files stay in the P2P cloud, as well as the synchronized devices, for as long as the user wishes | The files remain on the synchronized devices, unless the user deletes them | The pieces of the files stay in the P2P cloud, as well as the synchronized devices, for as long as the user wishes |
| Availability of files | Files are available for download at any time, as long as the cloud hosting them are up and running | Files are available for download at any time, as long as the cloud hosting them are up and running | Files are available for download at any time, as long as the cloud hosting them are up and running | Files are available for download at any time, as long as the cloud hosting them are up and running | Files are available for download at any time, as long as the nodes with the pieces of the files are online | Files are available for download as long as at least two nodes with the files to be synchronized are online | Files are available for download at any time, as long as the nodes with the pieces of the files are online |

\* Different regions have different prices; US has the cheapest price at $1.99/month

P2P-based systems offer potentially unlimited storage, as each node in the network acts as a server as well as a client. Furthermore, since the pieces of files are replicated on multiple nodes, even if a node is (or a set of nodes containing those pieces are) offline, downloaders can obtain those pieces from the online nodes, thus making the files more readily available for synchronization, and the network more reliable. All nodes in the network need to go offline at the same time for the network to be completely down. Moreover, since the pieces are encrypted, and scattered across the network, security and privacy are ensured in such systems.

Resilio Connect is the only P2P-based synchronization system that is powered by BitTorrent, and inherits almost all the benefits of other P2P-based systems. Although the table shows that Resilio Connect may not have the same level of performance and availability of files as the other systems, a BitTorrent powered synchronization can, in fact, be developed with these advantages.

In what follows, we discuss why we expect Resilio Connect, and more generally BitTorrent-based systems, to be the go-to technology for file synchronization systems.

### 4.1. BitTorrent Advantages

The main reasons in focusing on BitTorrent in this paper to give insights on the superiority of BitTorrent-powered, P2Pbased file sharing and synchronization systems are:

#### 4.1.1. Popularity

According to statistics released by BitTorrent, Inc. [24], there are 45 million daily active users, whereas on a monthly scale, a staggering 170 million users are active each month.

BitTorrent is very popular among the younger population, with 63% of the users aged 34 and below [24]. Furthermore, most of these users are "educated and tech-savvy" males, according to BitTorrent.

It should be noted that, although coming from the official website, these statistics are not complete, as it is rather difficult to collect stats on BitTorrent, due to its nature of being used in a decentralized, and at many times a private networking environment.

#### 4.1.2. Availability

Since BitTorrent is a P2P network, the complete file is almost always available to be downloaded, as long as a single peer is online in the network (assuming it contains the whole file). Furthermore, since the files are divided into pieces, individual pieces can be downloaded from the online nodes. Missing pieces can be downloaded from nodes once they come online. Comparing this to a client/server architecture, wherein a server holds the file to be downloaded, one can definitely see how reliable a P2P network is, more so the BitTorrent protocol.

#### 4.1.3. Performance

Several research works focused on the capabilities of a P2P network, many of which report the performance gains when downloading files using BitTorrent.

Raymond et al. measured the load on centralized servers when using BitTorrent conjointly [25]. The paper showed how BitTorrent reduces the load on a server, and increases the download performance. Using various technologies and measurements, this research presents various tests and analyses results on the performance of the BitTorrent protocol.

As noted above, along with the performance gains, BitTorrent, being a P2P protocol, also reduces the server load by making each node in the network act like a server. Moreover, the network adjusts accordingly to new nodes joining it, or nodes going offline, thus making the network more scalable.

### 4.1.4. Scalability

In a P2P system, each client is a potential server. That is, increasing demand translates into increasing offer. This results in the unique scalability that characterizes P2P systems. This is unlike a typical server/client architecture, in which a server has to handle an increase, or even a decrease in the number of connected clients. An increase in the number of clients increases the server load, whereas a decrease in the number makes the system less efficient.

### 4.2. BitTorrent Limitations

BitTorrent may inherit the advantages of a P2P network, but it does come with its limitations. The most prominent limitation of the protocol being is its security. There is a number of well-known security holes in BitTorrent [26], [27], including Authentication, Authorization and Trust & Reputation.

We reviewed the already available P2P file synchronization technologies that have already implemented security in their systems. One of the best examples of such a service is Symform, which encrypts file chunks before replicating them on the network [28]. These systems provide confidentiality and data integrity by encrypting the file chunks. They also provide authentication of the user, by a username and password combination, prior to sharing or downloading files.

## 5. CONCLUSIONS

We reviewed various file sharing and synchronization technologies and services in this paper. We also compared and discussed these technologies and services, and presented our arguments on why we believe that P2P-based, or more specifically, BitTorrent powered file synchronization systems are superior to traditional cloud-based file synchronization systems, and should be the go-to technologies for reliable and secure file sharing and synchronization services. Future works should focus on enabling online collaboration over P2P-based synchronization systems.

**Authors**

Zulqarnain Mehdi (Zul) is currently pursuing his MSc degree in IT (Software Systems) from Heriot-Watt University, Dubai. He is currently employed as a Software Engineer in a Dubai-based company.

Zul's research interests include cloud storage systems, file sharing, peer-to-peer, and BitTorrent.

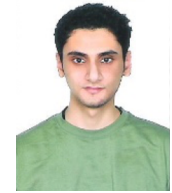

Hani RAGAB received the MSc degree from the university of technology of Compiegne (UTC), France, in 2003, and the Ph.D degree from the same university in 2007. He is currently a lecturer at Heriot-Watt University, United Kingdom.

His research interests include malware analysis, access control systems, peer-to-peer, and digital forensics.

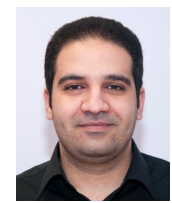